# Topological Electronics: From Infinity to Six

**Frank Z. Wang[1]**
[1]Division of Computing, Engineering & Mathematics Sciences, University of Kent, Canterbury, CT2 7NF, the UK
Correspondence to f,z,wang@kent.ac.uk. This research was partially conducted in the EC grant "Re-discover a periodic table of elementary circuit elements", PIIFGA2012332059, Marie Curie Fellow: Leon Chua (UC Berkeley), Scientist-in-Charge: Frank Wang (University of Kent).

*Abstract*— **Topology captures the essence of what remains unchanged under a transformation. This study was motivated by a newly found topological invariant called super conformality that leads to local activity of a higher-integral-order electric element. As a result, the traditional periodic table of the electric elements can be dramatically reduced to have only 6 passive ones (resistor, inductor, capacitor, memristor, meminductor, and memcapacitor), in contrast to the unbounded table predicted 40 years ago. Our claim was experimentally verified by the fact that the two higher-integral-order memristors in the famous Hodgkin-Huxley circuit are locally active with an internal battery.**

## I. Introduction: Topology in Computational Electronics

Based on the two fundamental attributes (electric charge $q$ and magnetic flux $\varphi$) [1][2][3][4], we found that all the (two-terminal) electric elements are topologically homeomorphic [5] in terms of physically interacting charge $q$ (or its $\alpha^{th}$-order variant $q^{(\alpha)}$) and flux $\varphi$ (or its $\beta^{th}$-order variant $\varphi^{(\beta)}$), in which $\alpha$ or $\beta$ can even be a fraction to reflect the fractional coupling between electricity and magnetism. Fig.1 describes such a topological homeomorphism, in which "genus" represents a class, kind, or group marked by common characteristics or by one common characteristic.

As a mathematical study of the structural properties of objects, topology is motivated by the fact that some scientific problems depend not on the exact geometric shape of the objects involved, but rather on the way they are organised and interconnected together [5]. As shown in Fig.1, a sphere and a cube have a property in common: both separate the space into two parts, the part inside and the part outside. Homeomorphism is the isomorphism in the category of topological spaces [5] — that is, two objects are homeomorphic if one can be continuously deformed into the other. Referring to the change in size or shape of an object, (continuous) deformation includes stretching and bending but excludes cutting or gluing.

In order to describe the physical charge-flux interaction in a electric element in Fig.2(b) and then predict the behavior of a real device, we used the LLG equation:

$$(1+g^2)\frac{d\overrightarrow{M_S}(t)}{dt} = -|\gamma|[\overrightarrow{M_S}(t) \times \vec{H}] - \frac{g|\gamma|}{M_S}[\overrightarrow{M_S}(t) \times (\overrightarrow{M_S}(t) \times \vec{H})],$$

where $M_S$ is the saturation magnetization, $H \propto i$ is a magnetic field along $z$, $g$ is the Gilbert damping parameter and $\varUpsilon$ is the gyromagnetic ratio [6][7]. The following expression was deduced:

$$m(t) = tanh\left[\frac{q(t)}{S_W} + C\right], \quad (1)$$

in which $m(t) = M_z(t)/M_S$ ($M_z$ is the Z component of $M_S$), $S_W$ is a switching coefficient and $C$ is a constant of integration such that $C = tanh^{-1}m_0$ ($m_0$ is the initial value of $m$) if $q(t=0)=0$ (no accumulation of charge at any point).

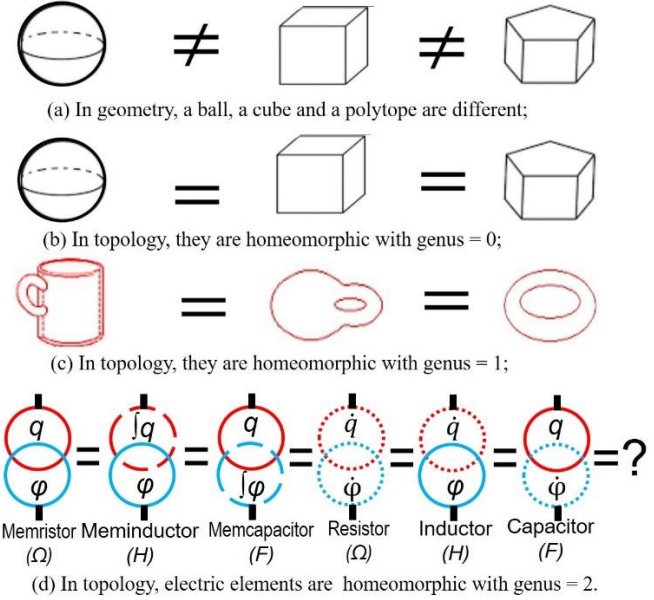

(a) In geometry, a ball, a cube and a polytope are different;
(b) In topology, they are homeomorphic with genus = 0;
(c) In topology, they are homeomorphic with genus = 1;

| Memristor | Meminductor | Memcapacitor | Resistor | Inductor | Capacitor |
| ($\Omega$) | (H) | (F) | ($\Omega$) | (H) | (F) |

(d) In topology, electric elements are homeomorphic with genus = 2.

*Fig.1 Topology is rooted in electric elements. We found that all the two-terminal electric elements are topologically homeomorphic with a genus of 2 (the number of holes is 2) in terms of physically interacting electric charge q (or its $\alpha^{th}$-order variant $q^{(\alpha)}$) and magnetic flux $\varphi$ (or its $\beta^{th}$-order variant $\varphi^{(\beta)}$). Intuitively, either electricity or magnetism can be represented by a circle in the sense that a moving charge (current) generates a circular magnetic field or a spin generates a circular magnetic flux [see Fig.2(b)]. As a quantity associated with topological space that do not change under continuous deformations of the space, a topological invariant is the number of holes in a surface (genus).*

By Faraday's law, the induced voltage $v(t)$ across the two terminals of the conductor is:

$$\mu_0 S \frac{dM_z}{dt} = S\frac{dB_z}{dt} = \frac{d\varphi_z}{dt} = -v(t), \quad (2)$$

where $\mu_0$ is the permeability of free space and $S$ is the cross-sectional area.

From Eq.2, we obtain

$$\varphi = \mu_0 SM + C' = \mu_0 SM_S m + C', \quad (3)$$

where $C'$ is another constant of integration.

Assuming $\varphi(t=0) = 0$, we have $C' = -\mu_0 SM_S m_0$, so

$$\varphi(q) = \mu_0 S M_s \left[\tanh\left(\frac{q}{S_W} + \tanh^{-1} m_0\right) - m_0\right]. \quad (4)$$

A typical $q$-$\varphi$ curve with $m_0 = -0.964$ is depicted in Fig.6(b), which agrees with those experimentally observed $q$-$\varphi$ curves [9][10][11]. By nature, it is nonlinear, continuously differentiable, and monotonically increasing (the three ideality criteria [2]). Hence, Eq.4 is reasonably used as one of the two exemplified constitutive curves for an ideal element $[q^{(\alpha)}, \varphi^{(\beta)}]$ that is defined based on the $q$-$\varphi$ interaction.

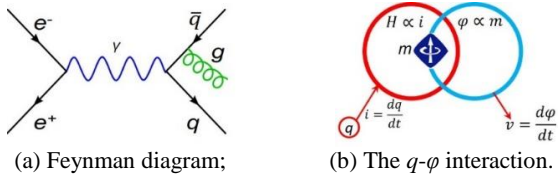

(a) Feynman diagram;            (b) The $q$-$\varphi$ interaction.

*Fig.2 (a) A Feynman diagram depicts vividly the interaction of subatomic particles such as electron ($e^-$), positron ($e^+$), photon ($\gamma$), quark ($q$), antiquark ($\bar{q}$), and gluon ($g$) [8]; (b) To give a similar visualization of what would otherwise be arcane and abstract formulas, a pair of entangled circles depicts the charge-flux interaction in a structure with a current-carrying conductor and a magnetic lump. The moving charge q (the current i) generates the magnetic field H that rotates the magnetization m inside the lump and consequently the switched flux $\varphi$ induces a voltage v across the conductor, thereby defining a generic "state-dependent Ohm's law" for all the (charge-controlled) two-terminal electric elements.*

The article is structured as follows. In Section I, we introduce topology and identify its potential applications in computational electronics. In Section II, we prove that the constitutive space of an electric element is a differential manifold. In Section III, we highlight that topology captures the essence of what remains unchanged and report that super conformality is such a topological invariant. In Section IV, we point out that the traditional periodic table of the electric elements can be dramatically reduced to have only 6 passive electronic elements based on our newly found super conformality. In Section V, we prove that a theorem that an electric element is locally active if its mid-point is non-zero. In Section VI, we mention an experimental verification in the famous Hodgkin-Huxley circuit and summarize our study.

## II. THE CONSTITUTIVE SPACE OF A ELECTRIC ELEMENT IS A DIFFERENTIAL MANIFOLD

Topologically, the constitutive $q$-$\varphi$ space of a electric element is a differential manifold (a type of manifold that is locally similar enough to a vector space to allow one to do calculus [12]) because it is globally defined with a differentiable but possibly complex structure (especially when characterising a higher-integral-order element).

As a topological space that locally resembles Euclidean space near each point, a manifold (with a possibly complex structure) is something similar to a globe (a 3D spherical model of the Earth), which cannot be unfolded onto a *2D* plane without distorting its surface due to the curvature. However, as shown in Fig.3, the surface of a globe can be approximated by a collection of 2D maps (also called charts), which together form an atlas of the globe. Although no individual map is sufficient to cover the entire surface of the globe, any place in the globe will be in at least one of the charts [12].

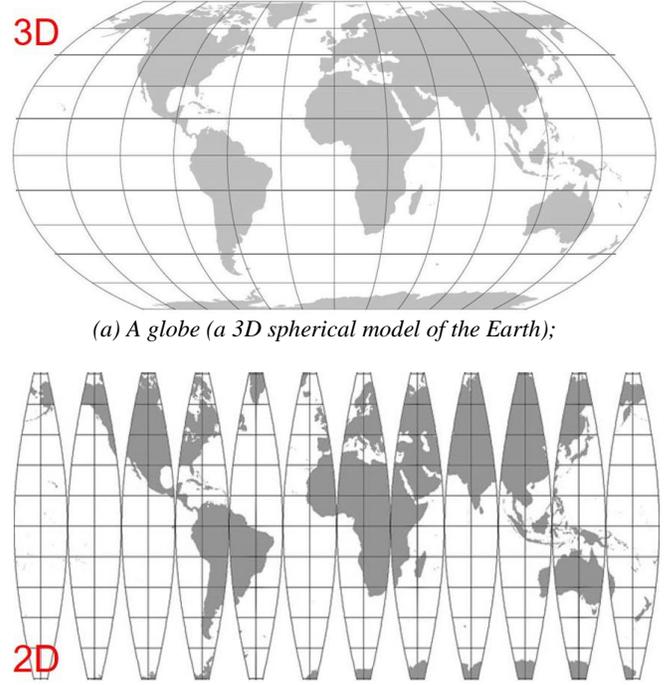

*(a) A globe (a 3D spherical model of the Earth);*

*(b) A collection of 2D maps (charts) forms the 3D globe.*
*Fig.3 A 3D globe can be approximated by a collection of 2D maps (charts).*

In topology, each map (chart) locally resembles a linear space near each point, as shown in Fig.4. These Euclidean pieces can be patched together to form the original manifold. A manifold can be described by two charts $U_1$ and $U_2$. If the transition ($f_1 \cdot f_2^{-1}$ or $f_2 \cdot f_1^{-1}$) from one chart to another is differentiable, then computations done in one chart are valid in any other differentiable chart.

Topologically, the constitutive space of an electric element is a differential manifold as shown in Fig.5. Some practical (nonideal) devices exhibit a double-valued $q$-$\varphi$ curve, and a pinched $i$-$v$ curve is asymmetric against the origin [13]. This is a generic case for an electric element in terms of physically interacting electric charge $q$ (or its $\alpha^{th}$-order variant $q^{(\alpha)}$) and magnetic flux $\varphi$ (or its $\beta^{th}$-order variant $\varphi^{(\beta)}$).

The mapping function ($f_1$ or $f_2$) can be the so-called conformal differential transformation when characterising a $0^{th}$-order element [1][2] or a more complex transformation [4] when characterising a higher-integral-order element.

Actually, the projection from the $q$-$\varphi$ plane to the $i$-$v$ plane in Fig.5 is based on the following theorem.

Theorem I: Conformality Theorem
The fractional differential transformation that covers both integer-order and fraction-order is conformal.

Proof:

$$\Psi = arctg\left(\frac{d^\Omega\hat{\varphi}(q)}{dq^\Omega}\right) = arctg\left(\frac{\frac{d^\Omega}{dt^\Omega}\hat{\varphi}(q)}{\frac{d^\Omega}{dt^\Omega}q(t)}\right) = arctg\left(\frac{v(t)}{i(t)}\right) = \Psi'$$

where $\frac{d^{-\Omega}}{dt^{-\Omega}}$ is the fraction-order calculus operator ($0 \leq \Omega \leq 1$).

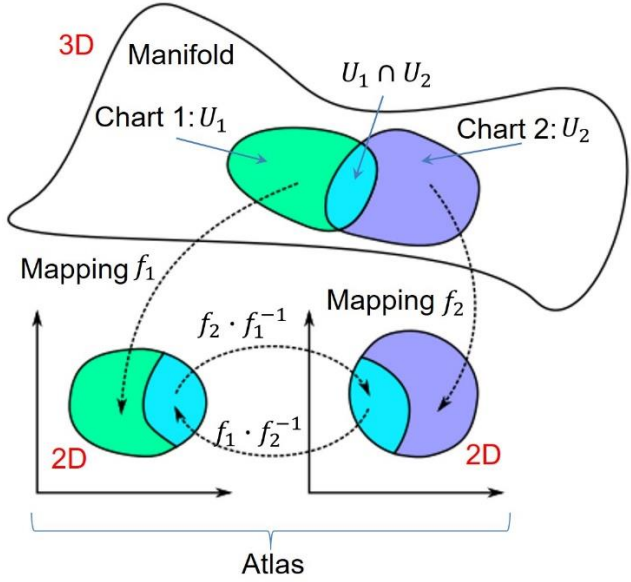

*Fig.4 Mapping $f_1$ or $f_2$ is a special type of relation in which one element of one domain is paired with another element of another domain. The transition map relates the coordinates defined by the two charts to one another: $f_1 \cdot f_2^{-1}$ and $f_2 \cdot f_1^{-1}$. To induce a global differential structure on the local coordinate systems induced by the homeomorphisms, their composition on chart intersections ($U_1 \cap U_2$) in the atlas must be differentiable functions on the corresponding linear space [12].*

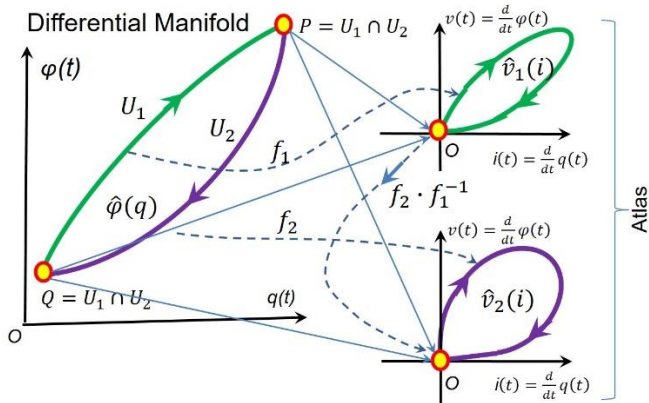

*Fig.5 Topologically, the constitutive space of an electric element is a differential manifold so that the two branches in the global structure can be mapped to the two local coordinate systems and analysed. Both points P and Q that join the two branches are projected to the origins of the two coordinate systems. Such a zero-crossing feature is a fingerprint of an ideal memristor [1][2]. A generic (nonideal) memristor has two q-φ characteristic branches and a pinched i-v hysteresis loop [13].*

The introduction of fraction-order calculus dealing with any order of derivatives or integrals [14][15] here describes some nonideal memristors whose charge-flux coupling is fractional. This theorem indicates that the differential transformation in the fraction-order still preserves angles in the same way as the (traditional) integer-order transformation does [1][2][4].

Obviously, the (differential) memristance of a point on the *q-φ* curve equals the chord memristance of the corresponding point on the *i-v* curve in the fractional transformation. This chord resistance actually defines a generic "state-dependent Ohm's law" for all the electric elements as follows:

$$v(t) = \frac{d^\Omega \hat{\varphi}^{(\beta)}(q)}{dq^{(\alpha)\Omega}} \cdot i(t) \tag{5}$$

Theorem II: Symmetry Theorem
An ideal memristor has an odd-symmetric voltage-current hysteresis loop and a symmetric memristance hysteresis loop if it is driven by an odd-symmetric periodic excitation current.

Proof:
If and only if the two branches ($U_1$ and $U_2$) in the *q-φ* plane (Fig.5) overlap and the two smooth mapping functions ($f_1$ and $f_2$) are the same, i.e., if and only if $U_1 = U_2$ and $f_1 = f_2$, the two *i-v* loops [$\hat{v}_1(i)$ and $\hat{v}_2(i)$] in the local coordinate systems will become identical:

$$\hat{v}_1(i) = f_1(U_1) = f_2(U_2) = \hat{v}_2(i) \tag{6}$$
$$\text{or } \hat{v}_1(i) = f_1 \cdot f_2^{-1}(\hat{v}_2(i)) = \hat{v}_2(i). \tag{7}$$

After rotating the 2nd coordinate system in Fig.5 by π, an odd-symmetric voltage-current hysteresis loop arises. Based on this theorem, an ideal electric element with memory should be characterised by a single-valued, unique and time-invariant constitutive curve complying with the following three criteria [2][13]: 1. Nonlinear; 2. Continuously differentiable; 3. Strictly monotonically increasing. Note that a curve is single-valued if it is strictly monotonic increasing. We need to unambiguously exclude the special case, in which a nonideal memristor has a doubled-valued, strictly-monotonic-increasing *q-φ* curve. Furthermore, the "single-valued" feature vividly depicts Noether's theorem [16] in the sense that every differentiable symmetry of the action of a physical system (that is the aforementioned odd-symmetry of the voltage-current hysteresis loop) has a corresponding conservative quantity (that is the number of the constitutive curve in this case).

In mathematics, conformality means the condition (of a map) of being conformal. Conventional conformality ($\psi = \psi', \gamma = \gamma', \eta = \eta', \theta = \theta'$) exists as a topological invariant under the conformal differential transformation from the constitutive (*q, φ*) plane to its first-order differential ($\dot{q}, \dot{\varphi}$) plane for an element.

From the concept of local passivity (the origin-crossing of the *v-i* loci) [17], the following local passivity theorem is reasoned:

Theorem III: Local Passivity Theorem
The first-order electric element is locally passive.

Proof:

Criteria 4 (strictly monotonically increasing) means that, if A<B, f(A)<f(B) by monotonicity, thus the slope of $\hat{\varphi}(q)$ is nonnegative ($f'(x_t) \geq 0$); hence, this ideal memrisor is locally passive at each point on the $\varphi$–$q$ curve. If the order "≤" in the definition of monotonicity is replaced by the strict order "<", then one obtains a strictly monotonically increasing function. Note that $f'(x_t) = 0$ only at those isolated points, rather than any continuous range, otherwise it violates the definition of monotonicity.

The origin-crossing signature leads to local passivity, which can be proved by contradiction: a cell is said to be locally active at a cell equilibrium point $\mathbb{Q}$ if, and only if, there exists a continuous input time function $i_\alpha(t) \in \mathbb{R}^m, t \geq 0$, such that at some finite time $T, 0 < T < \infty$, there is a net energy flowing out of the cell at $t = T$, assuming the cell has zero energy at t = 0, namely, $w(t) = \int_0^T v_\alpha(t) \cdot i_\alpha(t) dt < 0$, for all continuous input time functions $i_\alpha(t)$ and for all $T \geq 0$, where $v_\alpha(t)$ is a solution of the linearized cell state equation about $\mathbb{Q}$ with zero initial state $v_\alpha(t) = 0$ and $v_b(t) = 0$ [17]. If $i = 0, v \neq 0$ or vice versa, we should have an intersectional point crossing the $i = 0$ or the $v = 0$ axis, which implies that the $v$-$i$ curve must inevitably enter either the second or the fourth quadrant of the $v$-$i$ plane. Therefore, a cell with $i = 0, v \neq 0$ or vice versa must be locally active since $i(t)$ and $i(t)$ always have opposite signs in the second or the fourth quadrant of the $v$-$i$ plane hence $w(t) = \int_0^T v_\alpha(t) \cdot i_\alpha(t) dt < 0$.

### III. SUPER CONFORMALITY AS A TOPOLOGICAL INVARIANT

In this section, super conformality in the generic form $(x, y)$ with any possible shape of the excitation was proposed based on strict mathematical deduction and reasoning. In Fig.6, $(x, y)$ represent the two constitutive attributes for a electric element. Taking a $2^{nd}$-order memristor ($\alpha$=-2, $\beta$=-2) as an example, beyond the $1^{st}$-order setting, it requires double-time integrals of voltage and current, namely, $x = \int q dt = \iint i dt, \dot{x} = q = \int i\, dt, \ddot{x} = i$ and $y = \int \varphi dt = \iint v\, dt, \dot{y} = \varphi = \int v\, dt, \ddot{y} = v$.

A $2^{nd}$-order ideal meminductor [3] should be characterized by a time-invariant $\int q - \iint \varphi$ curve, thus $x = \int q\, dt, \dot{x} = q, \ddot{x} = i$ and $y = \iint \varphi dt, \dot{y} = \int \varphi dt, \ddot{y} = \varphi$.

A $2^{nd}$-order ideal memcapacitor [3] should be characterized by a time-invariant $\iint q - \int \varphi$ curve, thus $x = \iint q\, dt, \dot{x} = \int q dt, \ddot{x} = q$ and $y = \int \varphi\, dt, \dot{y} = \varphi = \int v\, dt, \ddot{y} = v$.

From $y_t = f_{x_t}(x_t)$ in the $x+jy$ plane in Fig.5, we have

$$\dot{z}_t = \dot{x}_t + j\dot{y}_t = \dot{x}_t + j\frac{d\,f(x_t)}{dt} = \dot{x}_t + j \cdot f'_{x_t}(x_t) \cdot \dot{x}_t. \quad (8)$$

Actually, Eq.8 verifies Theorem I (Conformality Theorem) between the $x+jy$ plane and the $\dot{x} + j\dot{y}$ plane (not shown in Fig.6): the line tangent $f'(x_t)$ to the $y_t = f(x_t)$ curve at any operating point $(x_t, y_t)$ in the $x+jy$ plane is equal to the (chord) slope of a straight line connecting the projected point $(\dot{x}_t, \dot{y}_t)$ to the origin in the $\dot{x} + j\dot{y}$ plane [1][2][4].

Observing Eq.8, we should obviously have $\dot{z}_t = \dot{x}_t + j\dot{y}_t = j \cdot f'_{x_t}(x_t) \cdot \dot{x}_t = j \cdot 0$ when $\dot{x}_t = 0$. That is, the curve in the $\dot{x} + j\dot{y}$ plane must go through the origin, as stated in Theorem III: (Local Passivity Theorem).

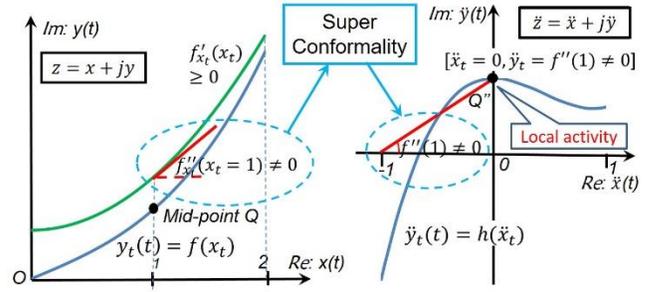

(a) The concave-convex orientation of the *x-y* curve is up;

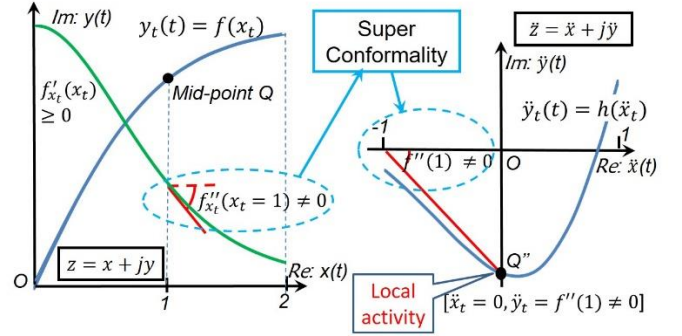

(b) The concave-convex orientation of the *x-y* curve is down.

*Fig.6 Super conformality was found to preserve a critical angle $f''_{x_t}(x_t = 1)$ between the constitutive x+jy complex plane and its second-order differential $\ddot{x} + j\ddot{y}$ complex plane. In this instance, "super" means that the conformality under the transformation from z=x+jy to $\ddot{z} = \ddot{x} + j\ddot{y}$ should be superior to the conventional conformality from x+jy to $\dot{x} + j\dot{y}$. This newly-found super conformality leads to local activity $\ddot{z}_t = \ddot{x}_t + j\ddot{y}_t = 0 + j\ddot{y}_t = j \cdot f''_{x_t}(x_t) \cdot \dot{x}_t^2 \neq j \cdot 0$ [17] of a $2^{nd}$—integral-order or higher-integral-order electric element. A polynomial $y = x + \frac{1}{3}x^3$ [2] and a logistic function described in Eq.4 were used in (a) and (b), respectively, to draw the graphs here without losing generality since the proof in the main text is carried out in the generic form (x, y). Modeling the exponential growth of a population, a logistic function represents a type of constitutive relation in terms of self-limiting [18]. The concave-convex orientation of the constitutive x-y curve determines the winding direction of the $\dot{x} - \dot{y}$ loop and the concave-convex orientation of the $\ddot{x} - \ddot{y}$ curve.*

Furthermore, we can use differentiation-by-parts to obtain

$$\ddot{y}_t = \frac{d\,\dot{y}_t}{dt} = f''_{x_t}(x_t) \cdot \dot{x}_t^2 + f'_{x_t}(x_t) \cdot \ddot{x}_t. \quad (9)$$

That is, when $\ddot{x}_t(t) = 0$, we obtain

$$\ddot{z}_t = \ddot{x}_t + j \cdot \ddot{y}_t = j \cdot f''_{x_t}(x_t) \cdot \dot{x}_t^2. \quad (10)$$

Eq.10 is a key finding of this topological electronics, which bridges the constitutive $x+jy$ complex plane and its second-order differential $\ddot{x} + j\ddot{y}$ complex plane.

Next, let us use proof-by-contradiction [19] to prove that $f''_{x_t}(x_t) \neq 0$. If $f''_{x_t}(x_t) = 0$ at every operating point, we should have $f'_{x_t}(x_t) = k$ where $k$ is an arbitrary constant and consequently $f_{x_t}(x_t) = kx_t + C$ where $C$ is another arbitrary constant. Obviously, $f_{x_t}(x_t) = kx_t + C$ is a linear function, which is in contradiction to Criterion 1 (nonlinearity) for an ideal electric element defined in Section II. Therefore, $\ddot{y}_t = f''_{x_t}(x_t) \cdot \dot{x}_t^2 \neq 0$ under any excitation $\dot{x}_t \neq 0$.

The following local activity [17] theorem for a 2nd-order or higher-integral-order electric element can then be obtained from Eq.10:

> **Theorem IV: Local Activity Theorem**
> A 2nd-integral-order or higher-integral-order electric element is locally active.

If we assume that an excitation $x_t(t) = 1 - \cos t$, we should have $\dot{x}_t(t) = \sin t = 1$ and $\ddot{x}_t = \cos t = 0$ when $t = \frac{\pi}{2}, \frac{3\pi}{2}, \ldots$. From Eq.10, we obtain the coordinate $\ddot{y}_t = f''_{x_t}(x_t) \cdot \dot{x}_t^2 = f''_{x_t}(x_t)$ when the second differential curve $\ddot{y}_t(t) = h(\ddot{x}_t)$ crosses the $y$ axis, i.e., $\ddot{x}_t(t) = \cos t = 0$. Therefore, the following super conformality theorem is reasoned from Eq.10:

> **Theorem V: Super Conformality Theorem**
> The second derivative $f''_{x_t}(x_t)$ at the mid-point $(x_t = 1, y_t)$ of the $y_t = f_{x_t}(x_t)$ curve in the $x+jy$ complex plane is equal to the slope of a straight line connecting the point $\ddot{z}_t = \ddot{x}_t + j\ddot{y}_t = -1$ to the point $\ddot{z}_t = \ddot{x}_t + j\ddot{y}_t = j \cdot f''_{x_t}(1)$ in the $\ddot{x} + j\ddot{y}$ complex plane.

This super conformality theorem is vividly depicted in Fig.6. A critical angle $f''_{x_t}(x_t = 1)$ is preserved between the constitutive $x+jy$ complex plane and its second-integral-order differential $\ddot{x} + j\ddot{y}$ complex plane.

## IV. A REDUCED PERIODIC TABLE OF ONLY SIX PASSIVE ELECTRIC ELEMENTS

Finding a correct, accurate electric element table in electrical/electronic engineering is similar to the discovery of Mendeleev's periodic table of chemical elements [20]. An electric element table would help understand the complex world of modern electronics and request rewriting the electrical/electronic engineering textbooks. Unfortunately, the previous unbounded table predicted 40 years ago [21][22] has an infinitive number of electric elements, as shown in Fig.7.

Based on our newly-found Theorem IV (Local Activity Theorem) and Theorem V (Super Conformality Theorem), a reduced periodic table (in green) of six electric elements is also displayed in Fig.7. In contrast, the Standard Model of elementary particles counts six flavours of quarks and six flavours of leptons [23]. Such a periodic table may help reveal the deep physical origin of elements, categorise the existing elements and predict new elements.

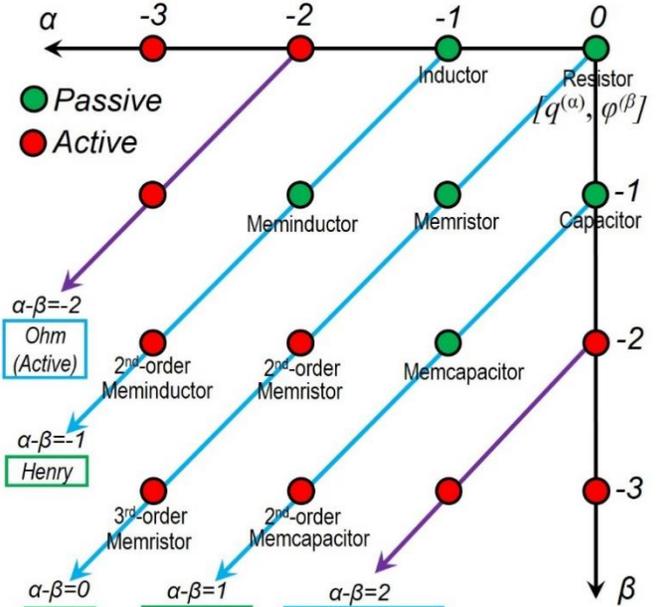

*Fig.7 Chua and Szeto predicted that most of the higher-integral-order electric elements $[q^{(\alpha)}, \varphi^{(\beta)}]$ are active and the only passive nonlinear candidates are on the three diagonals (in blue) [21][22]. The four-element torus (resistor, inductor, capacitor and memristor) was used as a seed to generate all other elements endlessly [1][2]. In this work, a reduced table (in green) was proposed to include only six electric elements: resistor, inductor, capacitor, memristor, meminductor, and memcapacitor. This new reduced table represents a big leap from an infinity in Chua's table [21][22] to a bound of 6 only.*

## V. MID-POINT THEOREM

Our study can be simplified to the following mid-point theorem when $x_t(t) = 1 - \cos t$:

> **Theorem VI: Mid-Point Theorem**
> If and only if the second-derivative at the mid-point of the constitutive curve is non-zero for a two-terminal electric element, it is locally active.

Actually, this theorem vividly describes our finding: all higher-integral-order passive memory electric elements [memristor (*α*≤-2, *β*≤-2), higher-integral-order meminductor (*α*≤-3, *β*≤-2), and higher-integral-order memcapacitor (*α*≤-2, *β*≤-3)] must have an internal power source (Fig.8).

According to the mid-point theorem, the passive version of a higher-integral-order electric element should not exist in nature. Even if it had existed, it would have degenerated into a zeroth-order negative nonlinear element. Nevertheless, the active version of a higher-integral-order electric element may still be found in nature (with either an electric, optical, chemical, nuclear or biological power source). The active version may also be built as an artifact in the lab with the aid of transistors, operational amplifiers, and/or power supplies.

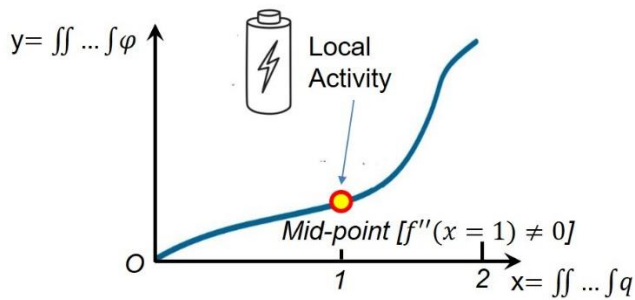

*Fig.8 Mid-Point Theorem. From Theorems IV and V, it can be reasoned that the mid-point of the constitutive curve for a two-terminal electric element is locally active [17].*

By coincidence, this phenomenon is quite similar to Mendeleev's periodic table of chemical elements, in which a chemical element with a higher atomic number is unstable and may decay radioactively into other chemical elements with a lower atomic number [20].

### VI. CONCLUSION & DISCUSSIONS

Topology is such a subject that examines geometric objects and captures the essence of what remains unchanged when they smoothly, continuously transform into one another. To date, topology has been used in many applications, such as biology, computer science, physics, robotics, games and puzzles, and fibre art. For example, the 2016 Nobel prize in physics was awarded for theoretical discoveries of topological phase transitions of matter [24].

Motivated by the above achievements, in this work, we used topology as a powerful tool to categorize the electric elements via a pair of integer/fraction numbers $(\alpha, \beta)$. When an electric element is described by the right pair of numbers $(\alpha, \beta)$, we found a brand new feature that is super conformality. As a result, a reduced periodic table of only six passive electric elements was then proposed, including resistor [25], inductor [26], capacitor [26], memristor [1], meminductor [3], and memcapacitor [3]. In our opinion, this reduced table reveals the topological homeomorphisms of the electric elements.

Our claim that all 2$^{nd}$-integral-orde or higher-integral-order electric elements must be active was experimentally verified by the fact that both higher-integral-order memristors [27][28] in the famous Hodgkin-Huxley circuit are locally active. Here, it is worth clarifying the relationship between an ideal electric element (normally in theory) and those real devices (in practice) belonging to the same family led by that ideal electric element [4]. An analogue in Chemistry is that Mendeleev predicted the existence of the (pure, ideal) chemical elements in his periodic table of chemical elements [20] and other chemists kept claiming they had found those elements one by one in spite of their purities are less than 100% in the real world. In 1971, to link flux $\varphi$ and charge $q$, Chua predicted an ideal memristor, whose state variable is $q$ or $\varphi$ [1]. 37 years later, HP announced "The missing memristor found" [29]. As a real device that is highly unlikely to be ideal, the HP memristor is not ideal as its state variable is the width "$w$" of the active T$_i$O$_2$ domain (doped with ions), which is still a function of charge $q$ (as well as other factors). Simply speaking, the state variable of the HP memristor is still $q$ (with the "purity" less than 100%). Despite its non-ideality, the HP memristor is still well-recognized as the first physical (1$^{st}$-order) memristor since it exhibits the zero-crossing signature as predicted [1]. That is, the 1$^{st}$-order memristor is passive. Similar to the physical HP memristor, both the 1$^{st}$-state-order potassium memristor and the 2$^{nd}$-state-order sodium memristor (in practice) are not ideal as their state variable are ionic gate probabilities, which are still a function of $q$ (with the "purity" less than 100%). The important thing is that these two higher-integral-order memristors are active with an internal battery [27][28], which agrees with our claim.

By coincidence, this dramatically-reduced number of electric elements with memory as key building blocks of modern electronics is analogous to a many-awards-winning breakthrough in mathematics: the bounded gap between two primes has been reduced from 70,000,000 to 6 [30]. Actually, two primes that differ by 6 are called sexy primes since "sex" is the Latin word for "six" [31]. In a similar way, this reduced periodic table of only six passive electric elements may also be called a sexy table.

**Funding**

This research was partially conducted in the EC grant "Rediscover a periodic table of elementary circuit elements", PIIFGA2012332059, Marie Curie Fellow: Leon Chua (UC Berkeley), Scientist-in-Charge: Frank Wang (University of Kent).

**Competing Interests**

The author has no relevant financial or non-financial interests to disclose.

**Author Contributions**

Frank Wang contributed to the study conception and design, data collection and analysis. Frank Wang wrote the manuscript.

**Data Availability**

Most of the datasets generated during and analysed during the current study are included in the manuscript. More datasets are available from the corresponding author on reasonable request.